\documentclass[%
 reprint,twocolumn,floatfix,
 amsmath,amssymb,
 aps,
 prl,
superscriptaddress
]{revtex4-1}
\usepackage{graphicx}
\usepackage{dcolumn}
\usepackage{bm}
\usepackage{comment}

\begin{document}

\title{Spin Seebeck effect and magnon diffusion length in $\rm{\mathbf{Fe}}_{\mathbf{3}}\rm{\mathbf{O}}_{\mathbf{4}}$}

\author{G. Venkat}
\affiliation{Dept. of Physics, Loughborough University, Loughborough LE11 3TU\\}

\author{C.D.W. Cox}
\affiliation{Dept. of Physics, Loughborough University, Loughborough LE11 3TU\\}

\author{D. Voneshen}
\affiliation{ISIS Neutron and Muon Source, Didcot, Oxfordshire, OX11 0QX}

\author{A.J. Caruana}
\affiliation{ISIS Neutron and Muon Source, Didcot, Oxfordshire, OX11 0QX}

\author{A. Piovano}
\affiliation{Institut Laue-Langevin, 6 rue Jules Horowitz, 38042 Grenoble Cedex 9, France}

\author{M.D. Cropper}
\affiliation{Dept. of Physics, Loughborough University, Loughborough LE11 3TU\\}

\author{K. Morrison}
\affiliation{Dept. of Physics, Loughborough University, Loughborough LE11 3TU\\}


\begin{abstract}

The determination of the magnon diffusion length (MDL) is important for increasing the efficiency of spin Seebeck effect (SSE) based devices utilising non-metallic magnets. We extract the MDL at $50$ and $300\,\rm{K}$ in an $\rm{Fe}_{3}\rm{O}_{4}$ single crystal from the magnon dispersion obtained using inelastic neutron scattering (INS) and find them to be equal within error. We then measure the heat flux normalised SSE responses and in-plane magnetization of $\rm{Fe}_{3}\rm{O}_{4}$ thin films and normalise by the static magnetization contribution to the SSE before determining the MDLs from a fit of the thickness dependence. We find that the MDLs determined in this way are smaller than that measured from INS which maybe due to differences in magnon propagation between bulk and thin film $\rm{Fe}_{3}\rm{O}_{4}$.

\end{abstract}

\maketitle

The spin Seebeck effect (SSE) was demonstrated in 2008 by Uchida {\em{et. al}} and has sparked a lot of interest for potential thermoelectric applications \cite{Ref1}. In the longitudinal configuration \cite{Ref2}, a spin current is generated along an applied temperature gradient which is normal to the magnetization. As spin currents cannot be detected easily, they are usually converted to charge currents in an adjacent metallic layer with high spin-orbit interaction (e.g. Pt) by the inverse spin Hall effect (ISHE) \cite{Ref2}. This discovery has paved the way for spin caloritronics that includes other thermomagnetic effects such as the spin Peltier and Nernst effects \cite{Ref3, Ref4}. Whilst, the efficiency of conventional thermoelectrics is limited by the interplay of thermal ($\kappa$) and electric ($\sigma$) conductivities \cite{Ref5}, the SSE is a candidate for improving the efficiency by allowing for the decoupling of $\kappa$ and $\sigma$ in the heavy metal and magnetic material \cite{Ref6}. The SSE is also a viable source of spin currents for spintronic applications \cite{Ref7}.

Although the SSE was initially demonstrated in a metallic ferromagnet \cite{Ref1}, it was later seen in the insulating ferrimagnet yttrium iron garnet (YIG) \cite{Ref8}, which has become the benchmark material for SSE studies \cite{Ref2}. There have also been studies of the SSE in half metals such as Heusler alloys \cite{Ref9,Ref10} and  $\rm{Fe}_{3}\rm{O}_{4}$ which are particularly relevant to spintronic applications due to the theoretically proposed 100\% spin polarisation at the Fermi level \cite{Ref11}. As the oldest known magnetic material, $\rm{Fe}_{3}\rm{O}_{4}$  has many technological applications \cite{Ref12} and has been attracting attention for SSE devices \cite{Ref13, Ref14, Ref15, Ref16}.

In an insulator the SSE is driven by magnon propagation \cite{Ref17} as opposed to being driven by spin polarized conduction electrons in metallic systems \cite{Ref3}. In the former case and for low conductivity magnets like $\rm{Fe}_{3}\rm{O}_{4}$, an important length scale for quantifying the maximum SSE efficiency is the magnon diffusion length (MDL) $\Lambda$ \cite{Ref18}. Quantitative knowledge of $\Lambda$ is also useful for studying fudamentally new physics such as magnon Bose-Einstein condensates \cite{Ref19}. Kehlberger {\em{et. al}} extracted $\Lambda_{\rm{YIG}}=90-140\,\rm{nm}$ from SSE measurements in YIG thin films, using an exponential model \cite{Ref20}. However, microwave measurements on YIG have found significantly larger MDLs $\sim400\,\rm{nm}$ \cite{Ref21}. In $\rm{MgO}/\rm{Fe}_{3}\rm{O}_{4}/\rm{Pt}$ thin films, the MDL was extracted from SSE measurements using the linear response theory (LRT) \cite{Ref22} and found to be $\Lambda_{\rm{Fe}_{3}\rm{O}_{4}}=17\,\rm{nm}$ at $300\,\rm{K}$ and increase to $40\,\rm{nm}$ at $70\,\rm{K}$ respectively \cite{Ref13}. Since $\rm{Fe}_{3}\rm{O}_{4}$ seems to show a temperature dependence of the MDL affecting the spintronic efficiency and there is some discrepancy in the MDLs obtained (for YIG), a careful study of the thermally driven spin current propagation in $\rm{Fe}_{3}\rm{O}_{4}$ is required. 

Another aspect of the SSE is its dependence on the saturation magnetization $M_{\rm{s}}$. A decrease in SSE in bulk single crystal YIG/Pt at $175\,\rm{K}$ has been correlated with an increase in magnetic surface anisotropy \cite{Ref23}. A similar decrease in the SSE in both bulk and thin films of YIG/Pt at $300\,\rm{K}$ at low magnetic fields has been attributed to the induced perpendicular magnetic anisotropy and its effect on the ISHE \cite{Ref24}. Thus, for comparing the SSE in different systems, we need to decouple the effects of the ISHE from the SSE and study the static magnetization and magnon propagation contributions to the SSE. 

In this Letter, we measure the magnon dispersion and group velocity in single crystal $\rm{Fe}_{3}\rm{O}_{4}$ using inelastic neutron scattering (INS). Subsequently, we fit integrated energy cuts of the magnon dispersion to simulated scattered neutron intensity profiles to obtain magnon linewidths and diffusion lengths at $300$ and $50\,\rm{K}$. We then report SSE measurements on $\rm{Fe}_{3}\rm{O}_{4}$ thin films at the same temperatures normalised to the heat flux $J_{\rm{Q}}=\frac{Q}{A}$ (where $Q$ is the heat passing through the sample and $A$ is the sample area) applied to the film. We also measure the variation of the in-plane magnetization with $\rm{Fe}_{3}\rm{O}_{4}$ thickness and show that it mirrors the trend of the SSE coefficient.  We proceed to normalise the SSE coefficient with the magnetization of the films to remove the static magnetization contribution to the measured SSE (introduced via the ISHE) and extract MDLs from the SSE measurements by fitting to the LRT, following the approach in \cite{Ref13}. We observe that the MDLs extracted from the SSE measurements are smaller than the ones obtained from INS which maybe due to differences in magnon propagation in bulk and thin film $\rm{Fe}_{3}\rm{O}_{4}$.

\begin{figure*}[hbt!]
\includegraphics[width=2\columnwidth]{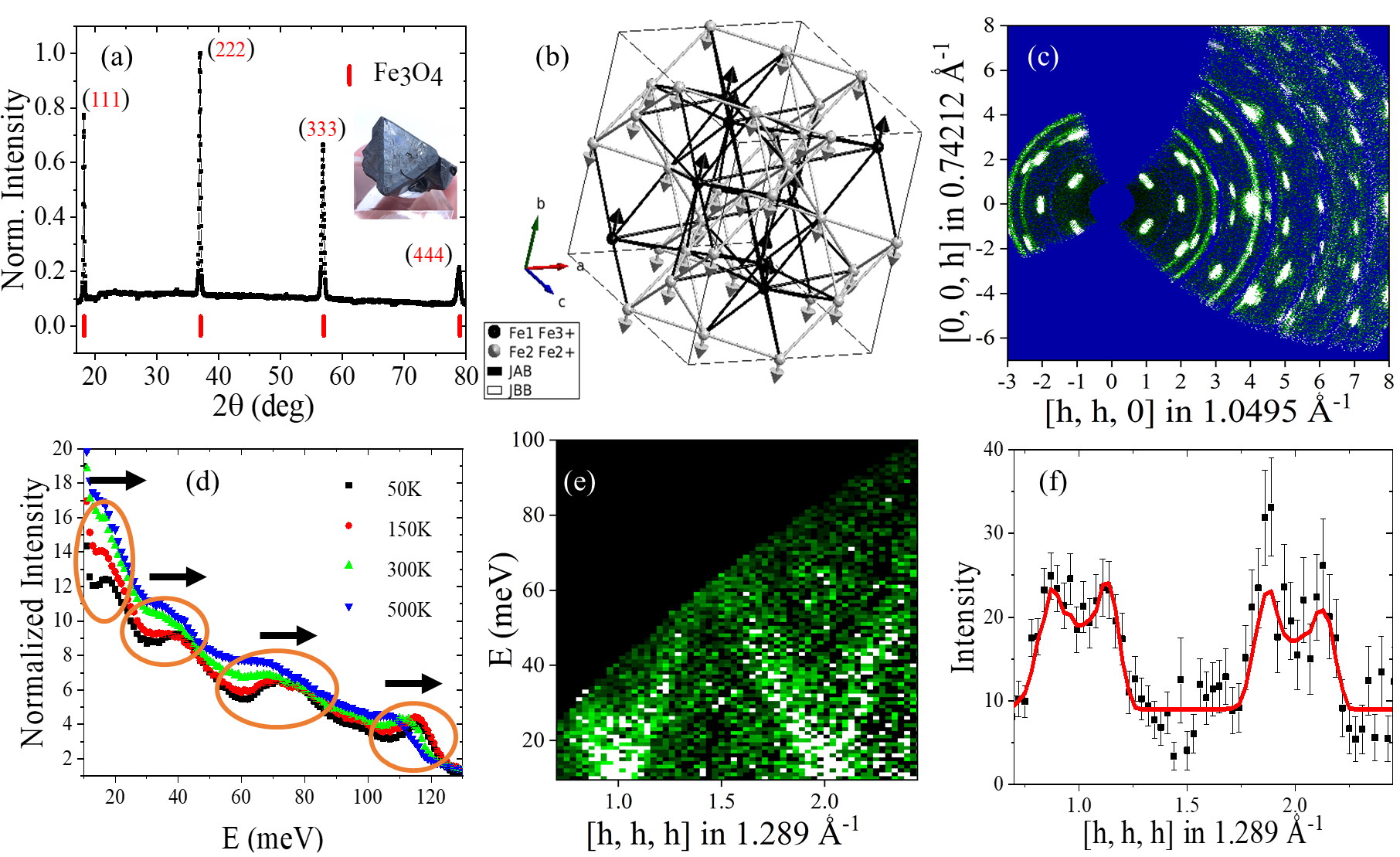}
\caption{(a) The XRD spectrum of the single crystal showing the [111] orientation. The red dashes mark the positions of the [111] (and harmonics) obtained from the ICSD database for $\rm{Fe}_{3}\rm{O}_{4}$ (ICSD: 26410). The inset shows a portion of the single crystal used for the measurement. (b) The magnetic structure of $\rm{Fe}_{3}\rm{O}_{4}$ showing the different magnetic sites, antiferromagnetic exchange $J_{\rm{AB}}$ (white lines) between sub-lattices and ferromagnetic exchange $J_{\rm{BB}}$ (black lines) within the B sub-lattice. (c) The scattered neutron Bragg peaks observed in the (HHH) plane corresponding to the magnetic structure. (d) The energy variation of scattered neutron intensity, integrated over all reciprocal space directions. The arrows indicate the blue shift of the magnon modes (marked by ellipses) with decreasing temperature (accompanied by narrowing of the spectra). (e) Magnon dispersion of $\rm{Fe}_{3}\rm{O}_{4}$ along the [HHH] direction at $300\,\rm{K}$. The almost linear variation of the dispersion is due to the antiferromagnetic exchange $J_{\rm{AB}}$ dominating in $\rm{Fe}_{3}\rm{O}_{4}$. (f) A line cut along $\mathbf{q}$ and integrated in energy from $20-30\,\rm{meV}$. This was fit (shown in red) to the simulated scattered neutron intensity for the $\rm{Fe}_{3}\rm{O}_{4}$ magnetic structure and used to extract the magnon linewidths.}
\label{fig:Fig1}
\end{figure*}

We used time-of-flight INS spectroscopy (on MAPS at the ISIS Neutron and Muon Source) to obtain the magnon dispersion at $50$, $150$, $300$ and $500\,\rm{K}$. The measurement was performed on a 5g naturally grown [111] oriented single crystal of $\rm{Fe}_{3}\rm{O}_{4}$ which was obtained from the Three Peaks mine in Utah, USA and is shown in the inset of Figure \ref{fig:Fig1} (a).  The XRD pattern in Figure \ref{fig:Fig1} (a) shows the [111] orientation of the crystal. It is worth noting that the magnon dispersion in $\rm{Fe}_{3}\rm{O}_{4}$ is isotropic along the [001], [110], and [111] directions \cite{Ref25} and so measurement along the [111] direction is representative. Figure \ref{fig:Fig1} (b) shows the cubic inverse spinel structure of $\rm{Fe}_{3}\rm{O}_{4}$ with two different symmetry sites for the Fe ions: A sites with tetrahedral coordination and occupied by $\rm{Fe}^{3+}$ ions and B sites with octahedral coordination and populated by a random distribution of $\rm{Fe}^{2+}$ and $\rm{Fe}^{3+}$ ions. The dominant magnetic interaction is the antiferromagnetic exchange coupling between the ferromagnetic A and B sub-lattices, making the overall magnetic structure ferrimagnetic. The magnetic exchange is well approximated by calculating the antiferromagnetic exchange constant $J_{\rm{AB}}$ and the ferromagnetic exchange constant of the B sublattice $J_{\rm{BB}}$ \cite{Ref26}. Figure \ref{fig:Fig1} (c) shows the Bragg peaks at $300\,\rm{K}$ obtained from INS corresponding to the magnetic structure. These peaks were used to correct the lattice  parameters and datasets for minor mis-alignments in the crystal orientation. 

Figure \ref{fig:Fig1} (d) shows the scattered neutron intensity integrated over all directions in reciprocal space, as a function of temperature. A pronounced blue shift and decrease in linewidth of the magnon modes with decrease in temperature can be seen which agrees well with the simulated magnon spectrum in other ferrimagnetic systems such as YIG \cite{Ref27}. Figure \ref{fig:Fig1} (e) shows the magnon dispersion along the [HHH] direction for the lowest order (acoustic) mode at $300\,\rm{K}$. The dispersion slope was used to extract a group velocity of $v_{\rm{g}}^{300\,\rm{K}}=13971\pm1819\,\rm{m/s}$ at $300\,\rm{K}$. This is in good agreement with the group velocity extracted from a complementary measurement using triple axis spectroscopy on the IN8 spectrometer at ILL Grenoble (more details are in the SI) from which we obtained $14631\pm800\,\rm{m/s}$.

Subsequently, four cuts along the $\mathbf{q}$ axis in Figure \ref{fig:Fig1} (e) and integrated along energy (from $20-30$, $30-40$, $40-50$ and $50-60\,\rm{meV}$) were simulatanously fit to the simulated scattered neutron intensity for the $\rm{Fe}_{3}\rm{O}_{4}$ magnetic structure (obtained using SpinW \cite{Ref28}). The fitting was done using the Tobyfit routine which accounts for instrument broadening in the linewidth \cite{Ref29}. The fits, shown for the $20-30\,\rm{meV}$ cut in Figure \ref{fig:Fig1} (f), gave an antiferromagnetic exchange constant of $J_{\rm{AB}}=-4.49\pm0.01\,\rm{meV}$ and a ferromagnetic exchange constant of $J_{\rm{BB}}=0.67\pm0.002\,\rm{meV}$ which agree well with literature \cite{Ref26}. We also extracted a linewidth of $\Gamma^{300\,\rm{K}} = 0.27\pm0.05\,\rm{meV}$ from the fits which corresponded to a magnon lifetime of $\tau^{300\,\rm{K}} = 2.44\pm0.45\,\rm{ps}$ at $300\,\rm{K}$. Using $v_{\rm{g}}^{300\,\rm{K}}$, we obtained a magnon diffusion length of $\Lambda_{\rm{Fe}_{3}\rm{O}_{4}}^{300\,\rm{K}} = v_{\rm{g}}^{300\,\rm{K}}\tau^{300\,\rm{K}} = 34.06 \pm 7.71\,\rm{nm}$. The magnon dispersion at $50\,\rm{K}$, gave us $\Lambda_{\rm{Fe}_{3}\rm{O}_{4}}^{50\,\rm{K}}=27.19 \pm 5.72$. We thus observe that values of $\Lambda_{\rm{Fe}_{3}\rm{O}_{4}}$ are equal within error at $50$ and $300\,\rm{K}$, with a small decrease observed at $50\,\rm{K}$ which might be due to decreased magnon population at this temperature. We regard these values obtained from bulk single crystal measurements as upper limits on the magnon diffusion length in $\rm{Fe}_{3}\rm{O}_{4}$.

\begin{figure*}[hbt!]
\includegraphics[width=2\columnwidth]{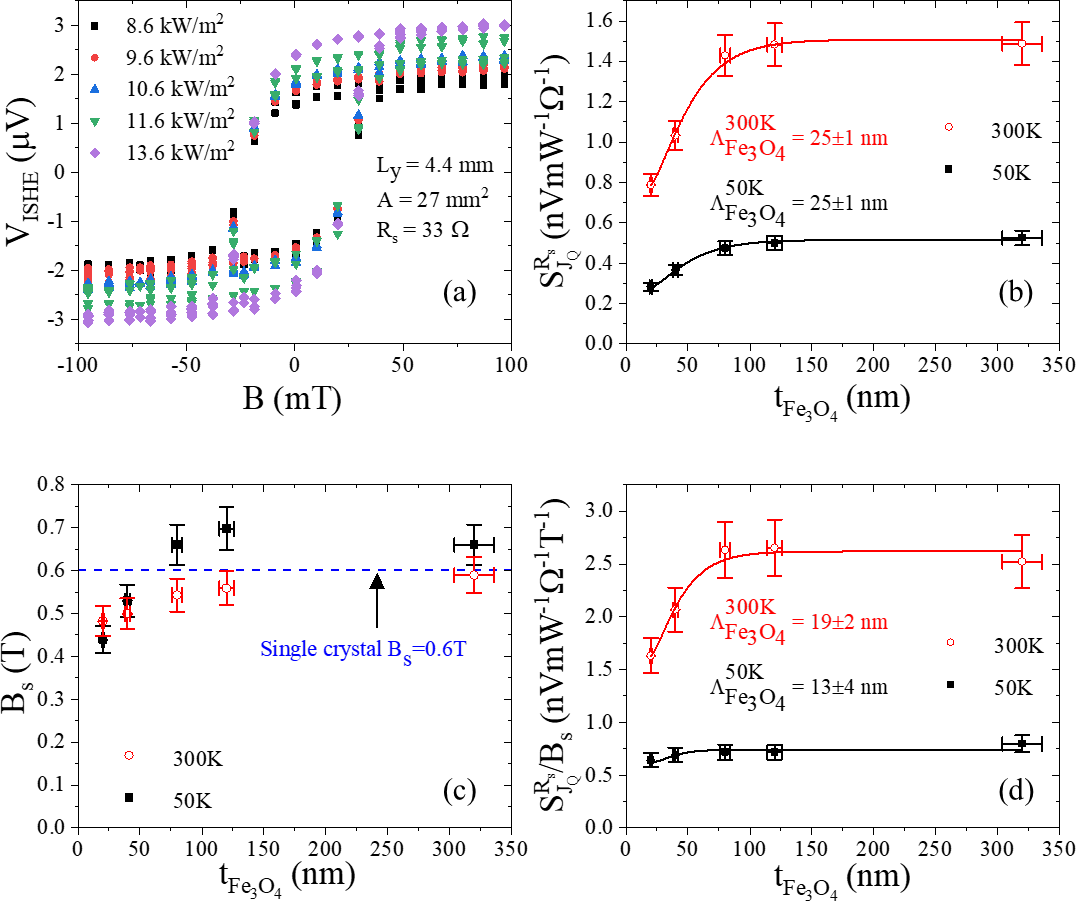}
\caption{(a) Example of the raw SSE voltage, V$_{\rm{ISHE}}$, as a function of the applied magnetic field for different heat fluxes $J_{Q}$ at $300\,\rm{K}$ for the $80\,\rm{nm}$ film. (b) Variation of the heat flux and sheet resistance normalised SSE coefficient $S_{\rm{J}_{\rm{Q}}}^{R_{\rm{s}}}$ with $t_{\rm{Fe}_{3}\rm{O}_{4}}$ at $50$ and $300\,\rm{K}$. The solid lines are fits obtained using eq. \ref{eq:3}. (c) The variation of the saturation magnetic flux density $B_{\rm{s}}$ with $t_{\rm{Fe}_{3}\rm{O}_{4}}$ at $50$ and $300\,\rm{K}$. The dashed line marks the single crystal value of $B_{\rm{s}}$ for $\rm{Fe}_{3}\rm{O}_{4}$. (d) Variation of $S_{\rm{J}_{\rm{Q}}}^{R_{\rm{s}}}/{B_{\rm{s}}}$ with $t_{\rm{Fe}_{3}\rm{O}_{4}}$ at $50$ and $300\,\rm{K}$. The solid lines are fits obtained using (\ref{eq:3}).}
\label{fig:Fig2}
\end{figure*}

We now describe the SSE measurements. The samples investigated in this study were pulsed laser deposition grown $5\times5\,\rm{mm}^{2}$ $\rm{SiO}_{2}/\rm{Fe}_{3}\rm{O}_{4}/\rm{Pt}$ films of thicknesses $t_{\rm{Fe}_{3}\rm{O}_{4}}=20,\,40,\,80,\,120\,\rm{and}\,320\,\rm{nm}$ and $t_{\rm{Pt}}=3-7.5\,\rm{nm}$ with both $\rm{Fe}_{3}\rm{O}_{4}$ and $\rm{Pt}$ growing with preferential [111] texture. More details of sample characterization can be found in the SI and details of the SSE measurement setup is part of a manuscript which is under preparation \cite{Ref40}. Figure \ref{fig:Fig2} (a) shows example SSE measurements for an $80\,\rm{nm}$ film, where the raw voltage, $V_{\rm{ISHE}}$, as the magnetic field $\mathbf{B}$ was varied is plotted for different applied (and measured) $J_{\rm{Q}}$. The increase in $V_{\rm{ISHE}}$ with $J_{\rm{Q}}$ is due to the generated spin current being proportional to the temperature gradient, $\nabla T$, across the $\rm{Fe}_{3}\rm{O}_{4}$ layer.

Since the SSE is usually associated with $\nabla T$, for bulk materials the SSE coefficient is often defined as \cite{Ref30, Ref31}
\begin{eqnarray}
\label{eq:1}
S_{\nabla T} &=& \frac{V_{\rm{ISHE}}L_{z}}{L_{y} \Delta T}\quad \frac{\rm{V}}{\rm{K}}, 
\end{eqnarray}
where $L_{y}$ is the contact separation and the $L_{z}$ is the sample stack thickness. However, for thin films, it was shown by Sola {\em{et. al}} \cite{Ref32} that the added complication of thermal resistance between the sample and the hot and cold baths makes $S_{\nabla T}$ unreliable \cite{Ref31}. They defined another coefficient \cite{Ref32}
\begin{eqnarray}
\label{eq:2}
S_{\rm{J}_{\rm{Q}}} &=& \frac{V_{\rm{ISHE}}}{L_{y} \rm{J}_{\rm{Q}}}\quad \frac{\rm{Vm}}{\rm{W}},
\end{eqnarray}
Due to the pitfalls associated with using $S_{\nabla T}$ as a comparative metric (\cite{Ref6,Ref31,Ref32}), we consider $\rm{J}_{\rm{Q}}$ normalisation to define the SSE coefficient in this paper.

An artefact contributing to the SSE is the anomalous Nernst effect (ANE), which is the thermal equivalent of the Anomalous Hall effect. The ANE has two possible contributions due to: (a) proximity magnetism induced in the Pt layer \cite{Ref33,Ref34,Ref35} and  (b) non-zero $\sigma$ of $\rm{Fe}_{3}\rm{O}_{4}$. Since the Pt thickness in our films was larger than $3\,\rm{nm}$, the proximity ANE was considered negligible \cite{Ref35}. The  heat flux normalised ANE coefficient of a bare $80\,\rm{nm}$ thick $\rm{SiO}_{2}/\rm{Fe}_{3}\rm{O}_{4}$ film was measured to be $15.9\pm2.8\,\rm{nVm}\rm{W}^{-1}$ at $300\,\rm{K}$. The resistance of the bare film was $756\,\rm{\Omega}$ while that of an $80\,\rm{nm}$ $\rm{Fe}_{3}\rm{O}_{4}$ /$5\,\rm{nm}$ Pt bilayer was $50\,\rm{\Omega}$. Using the approach in \cite{Ref13}, we obtained a corrected ANE response of $0.98\pm0.17\,\rm{nVm}\rm{W}^{-1}$ which is $<2.5\%$ of the SSE signal reported here. We therefore ignore the ANE in our calculations. This is supported by measurements of Anadon {\em{et. al}} who found a negligible ANE contribution to the SSE signal ($\sim7\%$) \cite{Ref13}. It is also worthwhile noting that the there is no ANE contribution from $\rm{Fe}_{3}\rm{O}_{4}$ at $50\,\rm{K}$ due to it being electrically insulating \cite{Ref36}.

In order to account for variation in the SSE signal due to minor variations in the thickness of the Pt layer, we further normalise $S_{\rm{J}_{\rm{Q}}}$ by the sheet resistance of the Pt layer $R_{\rm{s}}$ (measured using the 4-probe technique and given in the SI) and define $S_{\rm{J}_{\rm{Q}}}^{R_{\rm{s}}}=\frac{S_{\rm{J}_{\rm{Q}}}}{R_{\rm{s}}}$. The variation of $S_{\rm{J}_{\rm{Q}}}^{R_{\rm{s}}}$ with $t_{\rm{Fe}_{3}\rm{O}_{4}}$ is shown in Figure \ref{fig:Fig2} (b) at $50$ and $300\,\rm{K}$. The decrease in SSE coeffcient with temperature is consistent with previous reports \cite{Ref13}.

We fit the variation of $S_{\rm{J}_{\rm{Q}}}^{R_{\rm{s}}}$ with $t_{\rm{Fe}_{3}\rm{O}_{4}}$ to that predicted by the linear response theory given by \cite{Ref13, Ref21}
\begin{align}
\label{eq:3}
& S_{\rm{J}_{\rm{Q}}}^{R_{\rm{s}}} \propto  \frac{\left( 1-{\rm{sech}\left(\delta_{\rm{FM}} \right)} \right)\left( 1-{\rm{sech}\left(\delta_{\rm{NM}} \right)} \right)}{\left( \rm{tanh}\left( \delta_{\rm{NM}} \right)+F_{s} \right) \left( \rm{tanh}\left( \delta_{\rm{FM}} \right)+F_{m} \right)- G_{s}G_{m}},
\end{align}

where $\delta_{\rm{FM}}=\frac{t_{\rm{Fe}_{3}\rm{O}_{4}}}{\Lambda_{\rm{Fe}_{3}\rm{O}_{4}}}$, $\delta_{\rm{NM}}=\frac{t_{\rm{Pt}}}{\Lambda_{\rm{Pt}}}$, $\Lambda_{\rm{Pt}}$ is the spin diffusion length of Pt, $F_{s}$, $F_{m}$, $G_{s}$ and $G_{m}$ are material constants. The fits were performed for $F_{m}$ and $\Lambda_{\rm{Fe}_{3}\rm{O}_{4}}$ with  $F_{s}=G_{s}=5$, $G_{m}=1$, $t_{\rm{Pt}}=5\,\rm{nm}$ and $\Lambda_{\rm{Pt}}=7.7\,\rm{nm}$ \cite{Ref37} and are shown in Figure \ref{fig:Fig2} (b). We obtained $\Lambda_{\rm{Fe}_{3}\rm{O}_{4}}^{\,\rm{300K}}=25.15\pm1.45\,\rm{nm}$ and $\Lambda_{\rm{Fe}_{3}\rm{O}_{4}}^{\,\rm{50K}}=24.61\pm1.15\,\rm{nm}$, which are equal within error.

However, consideration of the in-plane magnetization should also be made due to its impact on the ISHE. The electric field generated by the ISHE is given by \cite{Ref24}, 
\begin{eqnarray}
\label{eq:4}
\mathbf{E}_{\rm{ISHE}} & \propto & \mathbf{J}_{\rm{s}}\times \mathbf{M},
\end{eqnarray}
where $\mathbf{J}_{\rm{s}}$ is the spin current generated along the direction of the applied heat flux which is along the thickness of the sample and $\mathbf{M}$ is the magnetization in the sample. We therefore measured the in-plane saturation magnetic flux density $B_{\rm{s}}=\mu_{0}M_{\rm{s}}$, at both $50$ and $300\,\rm{K}$ as a function of $t_{\rm{Fe}_{3}\rm{O}_{4}}$ (shown in Figure \ref{fig:Fig2} (c)). The single crystal value of $B_{\rm{s}}$ at both $50$ and $300\,\rm{K}$ is $\approx0.6\,\rm{T}$ \cite{Ref38} and we can see that for $t_{\rm{Fe}_{3}\rm{O}_{4}}>40\rm{nm}$, $B_{\rm{s}}$ for the thin films approach this value. We observe a decrease in $B_{\rm{s}}$ for lower $t_{\rm{Fe}_{3}\rm{O}_{4}}$ which is very similar to the variation of $S_{\rm{J}_{\rm{Q}}}^{R_{\rm{s}}}$ and expect that a decrease in $t_{\rm{Fe}_{3}\rm{O}_{4}}$ causes an out-of-plane canting of the magnetization and thus a decrease in the in-plane magnetization \cite{Ref39}. We also note that the decrease in $B_{\rm{s}}$ is more pronounced at $50\,\rm{K}$.

To disentangle the SSE from the static magnetization dependence introduced via the ISHE, we monitor the variation of $\frac{S_{\rm{J}_{\rm{Q}}}^{R_{\rm{s}}}}{B_{\rm{s}}}$ with $t_{\rm{Fe}_{3}\rm{O}_{4}}$ (shown in Figure \ref{fig:Fig2} (d)), which from Eq. (\ref{eq:4}) is proportional to $\mathbf{J}_{\rm{s}}$. We observe that while a decreasing trend for lower $t_{\rm{Fe}_{3}\rm{O}_{4}}$ is still evident at $300\,\rm{K}$, it is considerably reduced at $50\,\rm{K}$, indicating that there is a larger contribution from $B_{\rm{s}}$ to the SSE response at this temperature. We also fit these trends (of the variation of $\frac{S_{\rm{J}_{\rm{Q}}}^{R_{\rm{s}}}}{B_{\rm{s}}}$ with $t_{\rm{Fe}_{3}\rm{O}_{4}}$) to that predicted by Eq. (\ref{eq:3}) (also shown in Figure \ref{fig:Fig2} (d)) and obtained $\Lambda_{\rm{Fe}_{3}\rm{O}_{4}}^{\,\rm{300K}}=19.24\pm2.27\,\rm{nm}$ and $\Lambda_{\rm{Fe}_{3}\rm{O}_{4}}^{\,\rm{50K}}=12.56\pm3.97\,\rm{nm}$, which are also close and show the small decrease at $50\,\rm{K}$ seen in the values extracted from the INS measurements. We believe that these are accurate estimates of the MDLs in the $\rm{Fe}_{3}\rm{O}_{4}$ thin films after accounting for the effect of the decrease in the in-plane magnetization with decreasing film thickness.

The variation of $S_{\rm{J}_{\rm{Q}}}^{R_{\rm{s}}}$ with $t_{\rm{Fe}_{3}\rm{O}_{4}}$ has some contribution from $B_{\rm{s}}$ and accounting for this leads to a decrease in the MDLs estimated from the LRT. The decreased MDL values estimated from thin film SSE measurements and fitting to the LRT are around half the values estimated from bulk single crystal INS measurements. This could be a result of defects or due to the presence of a minor $\alpha$-Fe phase (seen in the XRD spectrum in Figure S5 in the SI) in the thin film, which can cause higher magnon scattering/damping and a lowering of the MDL. Nevertheless, INS measurements provide an upper limit on the expected MDL at $300$ and $50\,\rm{K}$. We also note that we find no evidence of the increase in MDL with decreasing temperature,which was reported in \cite{Ref13}, from both INS and SSE/LRT estimates. We expect that some of the trends observed in \cite{Ref13} are skewed by the uncertainties introduced by considering $S_{\nabla T}$ as the SSE coefficient.

In conclusion, we have measured the magnon dispersion in a single crystal of $\rm{Fe}_{3}\rm{O}_{4}$ using inelastic neutron scattering as a function of temperature and observe that the extracted MDLs ($\Lambda_{\rm{Fe}_{3}\rm{O}_{4}}^{300\,\rm{K}}=34.06 \pm 7.71\,\rm{nm}$ and $\Lambda_{\rm{Fe}_{3}\rm{O}_{4}}^{50\,\rm{K}}=27.19 \pm 5.72$). We also measured the SSE normalised to heat flux and saturation magnetic flux density in $\rm{SiO}_{2}/\rm{Fe}_{3}\rm{O}_{4}/\rm{Pt}$ thin films and normalise by the static magnetization contribution to the SSE (introduced via the ISHE) before determining the MDL ($\Lambda_{\rm{Fe}_{3}\rm{O}_{4}}^{\,\rm{300K}}=19.24\pm2.27\,\rm{nm}$ and $\Lambda_{\rm{Fe}_{3}\rm{O}_{4}}^{\,\rm{50K}}=12.56\pm3.97\,\rm{nm}$) from a fit of the thickness dependence. We find that the MDLs determined in this way are smaller than that measured from INS measurements. We hope that these studies will highlight the importance of disentangling the role of various effects in SSE measurements as well as motivate further studies to relate the SSE to the variation in $B_{\rm{s}}$ and MDL in magnetic thin films.

\begin{acknowledgments}
We would like to acknowledge the use of the facilities and the assistance of Keith Yendall in the Loughborough Materials Characterization Centre and Gavin Stenning in the Materials Characterization Laboratory at the Rutherford Appleton laboratory. We also acknowledge the contributions of A. Sola and V. Basso from INRIM, Italy who helped develop the low temperature heat flux SSE measurement setup. Beamline experiments at the ISIS Neutron and Muon Source were supported by a beamtime allocation from the Science and Technology Facilities Council. This work was supported by the EPSRC Fellowship (EP/P006221/1). All supporting data will be made available via the Loughborough data repository under doi 10.17028/rd.lboro.2001261. The MAPS data is available via doi 10.5286/isis.e.rb1820362.
\end{acknowledgments}
 
\bibliography{refs}
\bibliographystyle{apsrev4-1}

\end{document}